\documentclass{cargese}

\usepackage{epsf}
\usepackage{epsfig}

\title{Is The Enigmatic Source 3EG J1828+0142 a Galactic Microblazar ?}

\author{Y. Butt\num{1}, presenting on behalf of: D.~F. Torres\num{2,3}, G.~E. Romero\num{4,5}, \\ J.~M. Paredes\num{6}, M. Rib\'o\num{6}, J. Mart\'{\i}\num{ 7}, J. Combi\num{4}, B. Punsly\num{8}}

\address{\num{1}Harvard-Smithsonian Center for Astrophysics, Cambridge, MA, USA}
\address{\num{2}Department of Physics, Princeton University, Princeton, NJ, USA}
\address{\num{3}Lawrence Livermore National Laboratory, Livermore, CA, USA}
\address{\num{4}Instituto Argentino de Radioastronomia, Villa Elisa, B.A., Argentina}
\address{\num{5}presently at: CNRS/CEA/Saclay and Universit\'e de Paris, Paris, France}
\address{\num{6}Departament d'Astronomia i Meteorologia, Universitat de Barcelona, Spain} 
\address{\num{7}Departamento de F\'{\i}sica, Escuela Polit\'ecnica Superior, Universidad de Ja\'en, Spain}
\address{\num{8}4014 Emerald St., Torrance, CA, USA}

\abstract{
The unidentified EGRET source 3EG J1828+0142, located at
$(l,\;b)\approx(31.9,\;5.8)$ (Hartman et al. 1999), is particularly enigmatic
since it displays very high variability over timescales of months (Torres et
al. 2000, Tompkins 1999) yet no blazar has been found within the 95\%
confidence contour (CC) of the gamma-ray location error-box (Punsly et al
2000). It also has an unusually steep $\gamma$-ray spectral index of
$\Gamma=2.76\pm0.39$. A multifrequency study of this source has been initiated
and may reveal the presence of a novel type of compact galactic object: a
microblazar. X-ray data, especially those of high spatial resolution from
CHANDRA, would be  particularly useful in establishing such a connection.
}

\begin{document}

\maketitle

\section{Introduction}

The EGRET telescope onboard the Compton Gamma-Ray Observatory
(CGRO) surveyed the entire sky in the $0.1-20$ GeV $\gamma$-ray
energy range and discovered 271 sources, of which roughly
two-thirds still remain unidentified (Hartman et al. 1999). The
source 3EG J1828+0142, located at $(l,\;b)\approx(31.9,\;5.8)$
(Hartman et al. 1999), is particularly enigmatic since it displays
very high variability over timescales of months (Torres et al.
2000, Tompkins 1999). However, no blazar has yet been found within the 95\%
confidence contour (CC) of the gamma-ray location error-box
(Punsly et al 2000). 3EG J1828+0142's variability ranks 9$\sigma$
above the average variability of known gamma-ray pulsars, making
it the second most variable low-latitude non-transient
$\gamma$-ray source. It also has an unusually steep $\gamma$-ray
spectral index of $\Gamma=2.76\pm0.39$. A multifrequency study of
this source has been initiated and promises to reveal the presence
of a novel type of compact galactic object.\\
\section{Radio and X-rays}
We have studied the region of the 3EG J1828+0142 error-box using
radio data from the large-scale surveys by Haslam et al. (1981),
Reich \& Reich (1986) (see Combi et al. 2001) and small-scale VLA
observations from the NVSS Sky Survey by Condon et al. (1998). In
Fig. 1 we show a filtered map of the radio field at 1.4 GHz with
the probability location contours of the 3EG source superposed. A
large, shell-type structure can be clearly discerned. It is a weak
radio source (the integrated flux density is $18.2\pm 2.1$ Jy at
1.4 GHz, the non-thermal spectral index found for the radio
emission is $\alpha=-0.72\pm0.18$), resembling a typical SNR. Its
distance is $\sim 940$ pc making the radius of the remnant $\sim
30$ pc and its age $\sim 4\times 10^4$ yr, assuming standard
values for the particle density of the ISM and the original energy
release (Punsly et al. 2000). Since the $\gamma$-ray source is
located on the boundary of the remnant, the hypothetical compact
source should have a birth velocity of $\sim 700$ km s$ ^{-1}$, in
agreement with recent estimates (Lyne \& Lorimer 1994). Fig.~1 
shows the point-like radio sources detected at 1.4 GHz with
flux densities above 5 mJy. Most of them have no entry in any
current catalog. We have computed spectral indices when possible,
and upper limits for those sources not detected at
5 GHz in Condon et al.'s (1994) survey.\\
\begin{figure}[h!]
{\includegraphics[width=8cm]{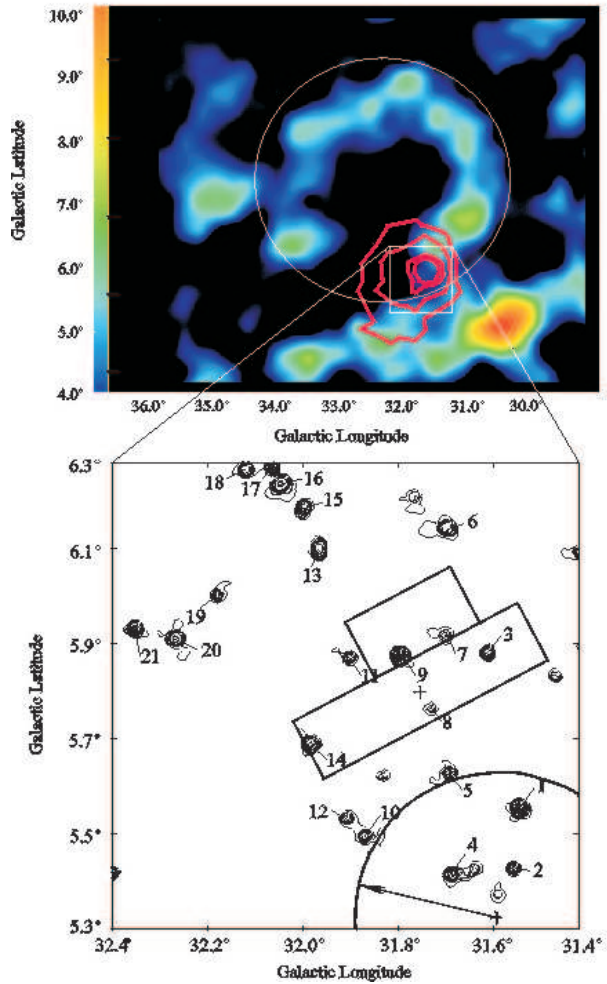}}%\hfill
\caption{{\small Upper panel: 1.4-GHz continuum map of the region
around 3EG J1828+0142.
%A Gaussian filtering beam of $90´\times
%90´$ has been applied to remove the diffuse background emission.
Radio contours are shown in steps of 0.01 K starting at 0.04 K.
The noise level is 0.02 K. Probability confidence contours (CC) for the 
location of the
$\gamma$-ray source have been superposed in red. Lower panel: Point-like
radio sources at the same frequency, detected by the VLA mostly
within the 95\% probability CC of the $\gamma$-ray source. The
selected sources are within the 68\% CC (Punsly et al. 2000). The
position of the requested ACIS-S field of view is shown. The part
of the circle in the lower-right corner was observed in pointed
mode by ROSAT, and we discarded their possible association with
the 3EG source due to poor spatial correlation. Sources \#12, 15 and 16, 
and possibly also 8, are
thermal. Sources \#13, 19, 20 and 21 are outside the 68\% CC,
whereas \#17 and 18 are outside even of the 95\% contour.}
\label{fig1}}
\end{figure}

%%%%%%%%%%%%%%%%%%%%%%\section{Archival X-ray Data}
The region of interest has been observed (but with only $\sim$340 sec
exposure) in the ROSAT All-Sky Survey (RASS), and only an small
corner (shown in the map)
%($\sim$0.1 deg$^{2}$ out of $\sim$2 deg$^{2}$ [[double
%check]]),
in pointed mode. We extracted all the ROSAT sources within 1
degree radius centered at $(l,b)=(31.8,5.9)$ using the standard
catalogs but none were found to be coincident with the radio
sources in Fig. 1. An ASCA observation which includes part of the
region of 3EG J1828+0142 is of too poor a spatial resolution to
determine conclusively any possible radio-X-ray correspondences.\\

\section{Modeling 3EG J1828+0142}

The rapid variability of 3EG J1828+0142 clearly points to a
compact object as the source of the $\gamma$-ray emission. But all
identified EGRET AGNs are also strong radio sources ($>$1Jy) with
flat spectra, as expected from synchrotron jet-like sources where
the $\gamma$-ray flux is the result of inverse Compton scattering.
No source in the radio field seems to correspond to a typical $\gamma$-ray
blazar. Independently, Mattox et al.~2001 also find it unlikely
(from an statistical point of view) that this source could be an
AGN. Although unexpected, discovering that the counterpart of the
3EG source is of extragalactic origin would be of great
importance also: it could signify the existence of a new class of strong
$\gamma$-ray emitting objects, which are radio quiet.

The galactic arena looks more promising, especially in light of the low Galactic
latitude of the source. Recently,
Kaufman-Bernado et al. (2002) have presented a model for
$\gamma$-ray emitting microblazars (MB) based on external Compton
scattering of thermal stellar photons from high-mass stellar
companions. Variability is naturally introduced by the Newtonian
precession of the accretion disk in the gravitational field of the
star. This is expected to result in the precession of the
relativistic jets, with the consequent flux variability due to
variable Doppler boosting. Depending on the parameters of the
model, such a MB can explain sources with a luminosity
$10^{34-38}$ erg/s of $\gamma$-rays in the observer's frame. The
non-thermal spectral energy distribution has two peaks, one from
the synchrotron jet emission (at radio-IR energies) and other from
Comptonization of stellar photons (at MeV-GeV energies). The total
X-ray emission is $\sim10^{34}$ erg s$^{-1}$. A model such as this
may also explain the LS 5039/ 3EG J1824-1514 connection recently
proposed by Paredes et al. (2000).  In general, if the jets are not
closely aligned to our line of sight, and particularly if the
binary companion is not of high mass, the X-ray spectrum of a MQ
will be affected by the disk emission. A power law plus an iron
line was the case for LS5039 (Rib\'o et al. 1999). There are no
known massive stars (spectral types B or O), superposed with any
of the radio sources shown in Fig. 1. However, this absence
could be explained by obscuration, since part of the region is in
the line of sight to Serpens-Cauda Molecular Cloud Complex, lying
at $\sim$250 pc from Earth (Straizys et al. 1996).

Another possibility for explaining 3EG J1828+0142 is an isolated black hole
(BH). It has been suggested that such objects could radiate
$\gamma$-rays due to spherical accretion from the interstellar
medium (Dermer 1997) or due to a charged magnetosphere (Punsly
1998, Punsly et al. 2000). The configuration of a simple
axisymmetric magnetosphere around a maximally rotating BH attains
a minimum energy configuration when the hole and the magnetosphere
have equal and opposite charge. Punsly (1998) has shown that the
magnetospheric charge can be supported in a stationary orbiting
ring or disk. The entire magnetized system is stable only in an
isolated environment, otherwise accretion onto the hole would
disrupt the ring and its fields. Magnetized black holes (MBHs) are
charged similarly to neutron stars in pulsars, but unlike them,
MBH have no solid surface and consequently no thermal X-ray
emission is expected. These objects can support strong magnetized
bipolar winds in the form of jets where $\gamma$-ray emission is
originated by IC emission with synchrotron seed photons (Punsly
1998). Since both magnetic and rotation axes are always aligned in
MBHs, their emission is non-pulsating (Punsly 1999). A detailed
MBH model of the $\gamma$-ray emission from 3EG J1828+0142,
consistent with its extreme variable behavior and steep spectral
index, was presented in Punsly et al. (2000). The broad-band SED
output of this model is shown in Fig. 2 of that paper. The
expected luminosity of the MBH in the CHANDRA energy range is
$\sim10^{34}$ erg s$^{-1}$, with a spectral index
$\Gamma=1.5\pm0.25$. In the CHANDRA range, the spectrum should
look like a featureless power law, there is no disk emission and
no lines are expected. The X-ray source should appear as a point
source positionally coincident with one of the non-thermal radio
sources detected in the field, since the same
leptonic population would also radiate synchrotron radio waves.\\
\section{Future Observations}

In order to locate any point-like X-ray emission coincident with
the sources in the radio map we have requested a modest amount of CHANDRA
time to explore the nature of sources 9, 11, and 14 in particular. These are 
the brightest non-thermal sources in the inner 68\% CC
of the gamma-ray source and are the most likely signatures of the putative
compact galactic object. (This region was not observed in pointed
mode by ROSAT.) CHANDRA's arcsecond spatial resolution will be extremely useful
for this work since it will at once allow us to conclude the
nature of the expected X-ray source (diffuse vs. point-like), and
will yield a sufficiently precise location to permit the follow-up
optical and VLBA studies which are being planned.
\\

% \section{Summary}
%
% \\

\noindent Combi J.A., Romero G.E., Benaglia P., Jonas J. 2001, A\&A 366, 1047.\\
Condon J.J., et al., 1994, AJ 107, 1829. \\
Condon J.J., et al., 1998, AJ 115, 1693.\\
%Douglas J.N., Bash F.N., Boyan F.A., 1996, AJ 111, 1945\\
 Dermer C.D., 1997, in Dermer C.D., Strickman M.S.,
Kurfess J.D. (eds.) Proceedings of the Fourth Compton Symposium,
AIP, New York, p.1275.\\
Hartman R.C., et al., 1999, ApJS 123, 79. \\
Haslam C.G.T., et al., 1981, A\&A 100, 209.\\
Kaufman Bernad\'o M.M., Romero G.E.,  Mirabel I.F., 2002, A\&A, 385, L10.\\
%Liu Q. Z., van Paradjis J., and van den Heuvel E. P. J. 2000,
%A\&AS 147, 25
Lyne A.G., Lorimer D.R., 1994, Nat 369, 127.\\
Mattox J. R., Hartman R. C., Reimer O., 2001, ApJS 135, 155.\\
%Mirabel, I. F., \& Rodr\'{\i}guez, L. F. 1999, ARA\&A, 37, 409.\\
Paredes J.M., Mart\'{\i} J., Rib\'{o} M., \& Massi M. 2000,
Sci, 288, 2340.\\
Punsly B., 1998, ApJ 498, 440.\\
Punsly B., 1999, ApJ 519, 336.\\
Punsly B., Romero G.E., Torres D.F., Combi J.A. 2000, A\&A, 364, 552.\\
Reich  W., Reich P., 1986, A\&AS 63, 205.\\
Rib\'o M., Reig P., Mart\'{\i} J., Paredes J. M., A\&A, 1999, 347,
518.\\
% Rib\'o, M., Paredes,J. M., Romero G. E., et al. 2002, A\&A, in press\\
 %Romero G.E., Benaglia P., Torres D.F., 1999,
 %A\&A 348, 868 \\ Romero, G. E. 2001, in The Nature of
 %Unindentified Galactic Gamma-Ray Sources, eds. A. Carraminana, O.
 %Reimer \& D.
 %Thompson, Kluwer Academic Publishers, Dordrecht, 65\\
Romero G.E., Kaufman-Bernado M.M.,  Combi J., and Torres D.F.,
2001 A\&A 376, 599.\\
Straizys V., Cernis K., and Bartsiute S. 1996, Baltic Astronomy 5,
124, \\
Tompkins W., 1999, PhD Thesis, Stanford University, Stanford, [astro-ph/0202141]. \\
Torres D.F., Romero G.E., Combi J.A., et al. 2001, A\&A,
370, 468.\\
Torres D.F., Romero G.E., Eiroa E. 2002, ApJ, In press,
[astro-ph/0112549].

%%%%%%%%%%%%%%%%%%%%%%%%%%%
%%%%% End of document %%%%%
%%%%%%%%%%%%%%%%%%%%%%%%%%%

\end{document}